\documentclass[rmp]{revtex4}

\usepackage{amsmath}
\usepackage{amscd}
\usepackage{amsfonts}
\usepackage{amssymb}
\usepackage{epsf}
\usepackage{revsymb}
\usepackage[dvips]{color}
\usepackage[cmtip,arrow]{xy}
\usepackage{pb-diagram,pb-xy}
\newtheorem{thm}{Theorem}%[section]
\newtheorem{defi}{Definition}%[section]
\newtheorem{lem}{Lemma}%[section]
\newtheorem{prop}{Proposition}
\newtheorem{cor}{Corollary}%[section]

\newtheorem{rem}{Remark}%[section]

\newcommand{\Lem}[2]{\begin{lem}\label{#1}#2\end{lem}}
\newcommand{\Thm}[3]{\begin{thm}[#1]\label{#2}#3\end{thm}}

\newcommand{\Cor}[2]{\begin{cor}\label{#1} #2\end{cor}}

\newcommand{\I}{\protect\ensuremath{\openone}}

\newcommand{\supp}[1]{\ensuremath{{\rm supp }(#1)}}

\newcommand{\tr}{\ensuremath{{\rm tr}}}

\newcommand{\id}{\ensuremath{{\rm id} \, }}

\newcommand{\Proof}[1] {\emph{Proof.} #1 \hfill $\blacksquare$ \vspace{16pt}} 
\def\1#1{{\bf #1}}
\def\2#1{{\mathcal #1}}
\def\4#1{{\tt #1}}
\def\5#1{{\sf #1}}
\def\6#1{{\mathfrak #1}}
\def\7#1{{\Bbb #1}}
\def\8#1{{\rm #1}}
\def\9#1{{\mathcurl #1}}

\DeclareFontFamily{OT1}{rsfs}{}
\DeclareFontShape{OT1}{rsfs}{m}{n}{<-7> rsfs5 <7-10> rsfs7 <10-> rsfs10}{}
\DeclareMathAlphabet\mathcurl{OT1}{rsfs}{m}{n}
\def\ic[#1]{{\tt [#1]}} %internal comment
\def\:{\mathpunct:}

\def\ic[#1]{{\tt [#1]}}

\def\pove#1#2{\9E(#1,#2)}
\def\regpove#1#2{\9R(#1,#2)}

\def\povm#1#2{\9M(#1,#2)}

\def\cbm#1#2{\9B(#1,#2)}

\def\d{\operatorname{d}}
\def\<{\langle}
\def\>{\rangle}

\begin{document}

\title{Barycentric decomposition of quantum measurements in finite dimensions}
\author{G.~Chiribella} %\inst{1}\inst{2} \and 
\author{G.~M.~D'Ariano} %\inst{1}\inst{2} \and 
\author{D.-M.~Schlingemann} %\inst{3}\inst{4}} 
\affiliation{Quantum Information Theory
Group, Dipartimento di Fisica A.~Volta, Universit\`{a} di Pavia,
via Bassi 6, 27100 Pavia, Italy} 
%\and 
\affiliation{INFN, Sezione di Pavia,  via Bassi 6, 27100 Pavia, Italy.}  
% \and 
\affiliation{ISI Foundation, Quantum Information Theory Unit,
 Viale S. Severo 65, 10133 Torino, Italy} %
\affiliation{Institut f\"ur
Mathematische Physik, Technische Universit\"at Braunschweig,
Mendelssohnstra{\ss}e~3, 38106 Braunschweig, Germany}
% Add name of the expert who has communicated your paper
%\communicated{name}
%
%\maketitle
%
\begin{abstract}
  We analyze the convex structure of the set of positive operator
  valued measures (POVMs) representing quantum measurements on a given
  finite dimensional quantum system, with outcomes in a given locally
  compact Hausdorff space. The extreme points of the convex set are
  operator valued measures concentrated on a finite set of $k \le d^2$
  points of the outcome space, $d<\infty$ being the dimension of the
  Hilbert space.  We prove that for second countable outcome spaces
  any POVM admits a Choquet representation as the barycenter of the
  set of extreme points with respect to a suitable probability
  measure. In the general case, Krein-Milman theorem is invoked to
  represent POVMs as barycenters of a certain set of POVMs
  concentrated on $k \le d^2$ points of the outcome space.
% according to a continuous
%  probability distribution, of POVMs concentrated on a finite number
%  of points.
\end{abstract}
\maketitle
%%%
\section{Introduction}
%%%
In the modern formalism of Quantum Mechanics the statistical
description of a measurement is provided by the concept of
\emph{positive operator valued measure (POVM)}
\cite{DaviesBook,HolevoBook,Helstrom,BuschBook}, whose introduction in
the literature on quantum probability dates back to the seminal papers
by Davies and Lewis \cite{DaviesLewis} and Holevo \cite{HolevoPOVM}.
A POVM associates to any possible event in a quantum experiment a
positive semidefinite operator on the Hilbert space of the measured
system, in such a way that the probability of the event is given by
the expectation value of the corresponding operator on the quantum
state describing the system preparation. The concept of POVM
generalizes, as far as it concerns the statistical aspects, the
traditional concept of ``observable'' by von Neumann \cite{VN}, which
turned out to be a too restrictive idealization to efficiently
describe actual experimental settings (such as the heterodyne
measurement in quantum optics \cite{hetero}), and even to give a
realistic modeling of photon-counting in the presence of losses
\cite{KK}.

In the case of finite dimensional quantum systems, the number of
different outcomes of a von Neumann observable must be finite, as the
number of eigenvalues of a self-adjoint operator cannot exceed the
dimension $d <\infty$ of the Hilbert space. Based on this observation,
it is commonly argued that all quantities measured on finite
dimensional systems must be intrinsically discrete or ``quantized''.
For example, when measured, a spin $j$ particle would be found in only
$d_j=2j+1$ possible spatial configurations, corresponding to the
possible values of the angular momentum along a given quantization
axis. The limitation on the number of possible values, however, only
holds for von Neumann measurements, which are a very particular subset
of all possible measurements in the statistical model of Quantum
Mechanics \cite{HolevoObservables}. If one considers arbitrary POVMs,
then there is no bound on the number of outcomes in an experiment, a
number which can be even uncountably infinite, despite the Hilbert
space dimension is finite.  This is indeed the case for the optimal
measurement of the spatial orientation of a spin $j$ that has been
devised in Ref. \cite{HolevoCovar}: in this measurement any direction
in the unit sphere is a possible outcome of the experiment.

From an operational point of view, a statement about the discreteness
of physical quantities of finite dimensional quantum systems cannot
rely on the concept of von Neumann observables. The question is then:
Is it possible to give a rigorous account to the intuitive idea that
the information carried by finite dimensional systems is intrinsically
discrete?  This intuitive idea is indeed supported by several
features, such as the existence of fundamental dimension-dependent
limits to the precision of phase measurements on atomic clocks
\cite{OptClocks}, to the extraction of directional information from
quantum gyroscopes \cite{refframe}, and to the maximum accessible
information in a coding-decoding scheme \cite{HolevoBound}.  Since all
mentioned limits arise in optimization problems where the goal is to
find quantum measurements that maximize some convex figure of merit,
it is natural to analyze the convex structure of the set of
measurements (POVMs) with given outcome space, expecting that the
discrete nature of information in finite dimensional systems will be
unveiled by the characterization of extreme points.

This paper fully characterizes the convex structure of the set of
POVMs with outcomes in a given locally compact Hausdorff space $Y$, by
\emph{i)} identifying the extreme points, and \emph{ii)} proving a
representation of arbitrary POVMs as barycenters of sets of POVMs with
finite outcomes.  We will first show that any extreme positive
operator valued measure is  concentrated on a finite number $k$ of
points, with $k$ not greater than $d^2$, the square of the Hilbert
space dimension. If $\Delta \subseteq Y$ is a possible event and
$M(\Delta)$ is the corresponding POVM operator, this means that an
extreme POVM $M$ must be of the form
\begin{equation}
M(\Delta) = \sum_{i=1}^k  \chi_{\Delta} (y_i)~ P_i~, 
\end{equation} 
where $\chi_{\Delta}$ is the indicator function of the set $\Delta$,
$\{y_i \in Y~|~ i=1, 2, \dots, k\}$ is a finite set of distinct
points, and $\{P_i~|~ i=1,2,\dots, k\}$ is a finite set of operators
forming an extreme POVM with finite outcome space $X=\{1,2, \dots,
k\}$, i.e.  $P_i\ge 0, \sum_i P_i=\openone_d$.  Operationally, this
means that any extreme POVM $P$ can be realized by first performing a
quantum measurement with finite set of outcomes $X=\{1,2,\dots, k\}$,
and then by injecting the result $i\in X$ in the outcome space $Y$ via
a post-processing rule $i \to y_i$.  This result reduces the
characterization of the extreme POVMs with locally compact outcome
space to the simpler characterization of extreme POVMs with finite
outcomes, which is has been extensively studied in the works by
St\"ormer \cite{Stoerm}, Parthasarathy \cite{Parthas}, and D'Ariano,
Lo Presti, and Perinotti \cite{ExtPaolino}.  Finally, we exploit
Choquet theorem to show that for second countable outcome spaces any
POVM can be represented as a barycenter of the set of extreme POVMs.
For general outcome spaces a barycentric representation in terms of
the closure of the set of extreme points is obtained instead by means
of Krein-Milman theorem.  In both cases, combining the barycentric
decomposition with the characterization of the extreme POVMs shows
that for finite dimensional quantum systems any measurement with a
continuum of outcomes is nothing but the randomized choice, according
to a continuous probability distribution, of a certain set of
measurements with finite outcomes.  In this sense, the continuum of
outcomes is simply equivalent to the presence of classical randomness
controlling the choice of the measuring apparatus. This provides the
rigorous and complete proof of the results presented in Ref.
\cite{ChiDArSchl07}.

%The barycentric decomposition presented here provides a rigorous
%operational statement of the quantization (i.e.  discretization) of
%physical properties of finite dimensional quantum systems.
%Differently from the quantisation arising from von Neumann
%observables, that does not allow more than $d$ outcomes, the
%quantisation arising from the barycentric decomposition allows
%outcomes up to $d^2$. This naturally agrees with the results of Ref.
%\cite{Latorre}, where it has been proved that the optimal measurement
%of the direction of a spin $j$ particle cannot be achieved by any von
%Neumann observable unless $j=1/2$: For $j>1/2$ the optimal finite
%measurement of direction has a number of outcomes $k$ with $d_j < k
%\le d_j^2$, $d_j= 2 j +1$ being the Hilbert space dimension.

It is worth stressing that all our results are derived for finite
dimensional Hilbert spaces, while in infinite dimensions the situation
is dramatically different. Indeed it is well known that von Neumann
observables always correspond to extreme POVMs, and any observable
with continuous spectrum is an example of extreme POVM with genuinely
uncountable outcome space, despite the Hilbert space has a
countable orthonormal basis. Moreover, a remarkable feature in
infinite dimensions is that von Neumann observables are dense in the
set of POVMs with given outcome space \cite{Holevodense}.

The paper is organized as follows: In Section~\ref{sec-1} we provide
the basic notation and definitions. In particular, we highlight the
equivalence between POVMs and regular \emph{operator valued
  expectations (OVEs)}, a class of positive maps that will be
extensively used in the statement and in the derivation of the main
results.  Regular operator valued expectations coincide with what is
known as \emph{quantization maps} in the literature on geometric
quantization \cite{ali,landsman}, namely positive maps from functions
on a classical phase-space to operators on the system's Hilbert space.
It is worth stressing that the present paper can be read as well as a
characterization of the extreme quantization maps for finite
dimensional quantum systems, along with a barycentric representation
of arbitrary quantization maps.  The characterization of extreme
POVMs/regular OVEs is carried out in Sec.  \ref{sec-2}.
Section~\ref{sec-3} presents a few topological properties that will be
useful for deriving barycentric decompositions. Finally, Section
\ref{sec-4} is devoted to the proof of barycentric representations of
POVMs and regular OVEs, first in the case of second countable outcome
spaces, and then in the general case.

\section{Positive operator valued measures and expectations}
\label{sec-1}

\subsection{Positive operator valued measures}

% The notion of positive operator valued measure originates from measure theory where the values of the measure are allowed to be positive operators instead of positive numbers only \cite{berberian}.

In the following $\8M_d$ and $\8M_d^*$ will denote the C*-algebra of
$d \times d$ complex matrices and the Banach space of linear
functionals on $\8M_d$, respectively.

\begin{defi} Let $Y$ be a measure space with $\sigma-$algebra $\sigma
  (Y)$. A \emph{positive operator valued measure (POVM)} in dimension
  $d < \infty$ is a map $M\:\sigma (Y) \to\8M_{d}$ that assigns to
  each measurable set $\Delta\in\sigma (Y)$ an operator $M(\Delta)\in
  \8M_d$ satisfying the following conditions:

\vspace{1em}

\noindent
{\bfseries Positivity:}  $M(\Delta)\geq 0 \quad \forall \Delta \in \sigma (Y)$ 

\vspace{1em}

\noindent
{\bfseries Normalization:} 
$M(Y)=\I_d$, with $\I_d \in \8M_d$ the identity matrix.

\vspace{1em}

\noindent
{\bfseries $\sigma$-Additivity:}
$M(\cup_{i\in\7N}\Delta_i)=\sum_{i\in\7N}M(\Delta_i)$ for any
countable family of mutually disjoint sets $\{\Delta_i \in \sigma (Y)
|i \in \mathbb N\}$, where the series converges weakly.

\end{defi}
%POVMs in dimension $d$ with given space $Y$ form a convex set, namely if $M_1$ and $M_2$ are POVMs then $pM_1 + (1-p) M_2$ is also a POVM for any $p \in [0,1]$. Such a convex  set will be denoted as $ \9P(Y,d)$. 

Throughout this paper the measure space $Y$ will be always a locally
compact Hausdorff space, and $\sigma (Y)$ will always denote the
Borel $\sigma-$algebra. The term POVM will be used as
a synonymous of \emph{regular} Borel POVM, as defined in the following:
\begin{defi} Let $Y$ be a locally compact Hausdorff space with Borel $\sigma-$algebra $\sigma (Y)$. A Borel POVM $M$ is called \emph{regular} if the condition
\begin{equation}
  M(\Delta) = \sup \{M (K)  ~|~ K \subseteq \Delta~,~ K ~ {\rm compact} \}
\end{equation}  
\vspace{1em} is fulfilled for any Borel set $\Delta \in \sigma (Y)$.
\end{defi}

The set of regular Borel POVMs is a convex set, denoted by $\povm Y
d$, and will be the focus of our investigation.

In quantum mechanics, any POVM yields the probabilities of events
occurring in a particular experimental setup. The elements of the
space $Y$ are the possible outcomes of the experiment, and $Y$ is
accordingly referred to as \emph{outcome space}. The possible events
are measurable subsets of $Y$, the subset $\Delta$ corresponding to
the event "the outcome of the experiment belongs to $\Delta$".  The
 states of a quantum system with finite dimensional Hilbert
space $\2H \simeq \mathbb C^d$ are positive normalized functionals
over the C*-algebra of complex matrices $\8M_d$.  For a quantum
system prepared in the state $\rho  \in \8M_d^*$ the probability of the event $\Delta$ is given by the Born rule
\begin{equation}
  p (\Delta)= \rho(M(\Delta))~.
\end{equation}  Accordingly, the
POVM $M$ assigns to every quantum state $\rho$ a classical
probability distribution $m_{\rho}$ via the relation $m_{\rho} (\Delta) = \rho(M(\Delta))$. Any bounded measurable function $f $ can
by averaged with respect to $m_{\rho}$, thus yielding
the expectation value
\begin{equation}\label{ExpVal}
  \mathbb E_{m_{\rho}}(f)=\int_Y m_{\rho} (\8dy)\ f(y) = \int_Y \rho (M(\d y))~ f(y) \; .
\end{equation}
The expectation $\mathbb E_{m_{\rho}}(f)$ in Eq. (\ref{ExpVal}) can be
extended by linearity to a unique functional on $\8M_d^*$, i.e. to a
unique operator $E (f) \in \8M_d$ satisfying the relation:
\begin{equation}\label{T(f)}
\rho(E(f))= \mathbb E_{m_{\rho}} (f)  \qquad \forall \rho \in \8M_d^*~.
\end{equation}
The map $E: f \mapsto E (f)$ can be viewed as an {\em operator valued
 expectation}: indeed, comparing Eqs. (\ref{ExpVal}) and (\ref{T(f)})
we obtain
\begin{equation}\label{conclu}
E (f)=\int M(\8dy)\ f(y) \; ,
\end{equation}
the integral converging in the weak operator topology \cite{berberian}.

\subsection{Operator valued expectations}
Dealing with locally compact Hausdorff spaces, it is convenient to
focus our attention to the C*-algebra $\9C_0 (Y)$ of continuous
functions vanishing at infinity, equipped with the $\sup$-norm $\|f\|=
\sup_{y \in Y} |f(y)|$.  In the following, we will consider $\9C_0
(Y)$ as a subalgebra of the unital C*-algebra of functions that are constant at infinity 
\begin{equation}
\begin{split}
  \overline {\9C_0} (Y) &= \9C_0 (Y) \oplus \7C\\
&= \{a f + b \I_Y ~|~ f \in \9C_0 (Y), \ a, b \in \mathbb C  \}~,
\end{split}
\end{equation} 
where $\I_Y$ is the constant function $\I_Y (y) =1 ~\forall y \in Y$.
Moreover, we will extensively use that fact that the C*-algebra $\overline
{\9C_0 }(Y)$, obtained by adding the unit to $\9C_0 (Y)$, is naturally
isomorphic to $\9C (\bar Y)$, the C*-algebra of continuous functions
on the one-point compactification $\bar Y = Y \cup \{\infty\}$ \cite{Wegge-Olsen}.

\begin{defi} An {\em operator valued expectation (OVE)} in dimension
  $d <\infty$ is a map $E: \overline{\9C_0} (Y) \to \8M_d$ that assigns to any
  function $f \in \overline {\9C_0} (Y)$ an operator $E(f) \in \8M_d$ satisfying
  the following conditions: \vspace{1em}

\noindent
{\bfseries Positivity:}  $E(f)\geq 0 \quad \forall f \ge 0$ 

\vspace{1em}

\noindent
{\bfseries Normalization:} 
$E(\I_Y)=\I_d$.

\end{defi} 

Operator valued expectations  form a convex subset of the set $\cbm
{Y}{d}$ of bounded maps from $\overline {\9C_0} (Y)$ to $\8M_d$, where the norm is
defined by 
\begin{equation}
\| E\| = \sup_{f \in \overline {\9C}_0 (Y): \|f\|=1} \| E(f)\|~,
\end{equation}
$\| O \|$ denoting the operator norm of $O \in \8M_d$. The set of all
operator valued expectations will be denoted by $\pove{Y}{d}$.

\begin{rem}\label{rem:unitball}  {\em 
Since the domain of the positive map $E \in \pove Y d$ is the abelian algebra $\overline{\9C_0}(Y)$,  $E$ is automatically completely positive
\cite{Paul86}. Therefore, for any OVE $E \in \pove Y d$ we have 
\begin{equation}\label{intheball}
\| E\|
= \sup_{0 \le f \le \I_{Y}} \| E(f)\| = \|E(\I_Y)\| =\|\I_d\|=1~.
\end{equation}
This shows that the set $\pove Y d$ is contained in the intersection
between the cone of positive maps and the unit ball in $\cbm Y d$.
Notice that such an intersection also contains positive maps that are
not OVEs: not any positive map $E$ with $\|E\|=1$ satisfies $E(\I_Y) =
\I_d$. }
\end{rem}

\begin{rem}{\em Since the unital algebra $\overline{\9C}_0 (Y)$ can be
    identified with $\9C(\bar Y)$, the set of OVEs $\pove Y d$ can be
    identified with the set of OVEs $\pove {\bar Y} d$, namely $\pove Y d \simeq \pove {\bar Y} d$. In  the following we will make often exploit this identification.}
\end{rem}
\subsection{Relation between POVMs and OVEs}

Each POVM $M \in \povm Y d$ induces an OVE $E \in \pove Y d$ via the relation (\ref{conclu}).  The
converse, however, is not straightforward, as in the definition of
OVE there are no requirements entailing  $\sigma-$additivity and regularity of
measures.  
%For non-compact outcome spaces there exist OVEs that do
%not correspond to any POVM.  
This motivates the following definition:
\begin{defi}\label{defi-regpove} An OVE $E\in\pove{Y}{d}$ is called
  {\em regular} if
\begin{equation}\label{eq-regpove}
\sup\{E(f)|f\in\9C_0(Y), 0\leq f\leq 1\}=\I_d
\end{equation} 

\end{defi}
The subset of regular OVEs will be denoted by $\regpove{Y}{d}$. Notice
that for compact outcome spaces $Y$ all OVEs are regular, namely
$\regpove {Y}{d} \equiv \pove {Y}{d}$.

As already mentioned in the introduction, regular OVEs are also known
as quantization maps in the literature on geometric quantization
\cite{ali,landsman}.  The relation between between regular OVEs
(quantization maps) and POVMs is a well known fact in such a
literature (see e.g. \cite{landsman}), and is reported here for completeness of presentation.

\Thm{Characterization of regular OVEs}{thm-riesz}{Let $Y$ be a locally
  compact Hausdorff space.  An OVE $E \in \pove Y d$ is regular if and
  only if there exists a POVM $M_E \in \9M(Y,d)$ such that
\begin{equation}\label{pove-povm}
E(f)=\int M_E(\8dy) \ f(y) \; .
\end{equation}
The above equation sets a one-to-one affine correspondence between $\regpove Y d $ and $\9M (Y, d)$.}  

\Proof{Let $E$ be an OVE.  Then for any state $\rho\in\8M_d^*$ the
  composition $\rho \circ E$ defines a state on $\overline{\9C}_0(Y)$.
  Moreover, $E$ is regular if and only if the restriction of
  $\rho\circ E$ to the ideal $\9C_0(Y)$ satisfies $\|\rho\circ
  E|_{\9C_0(Y)}\|=1$, namely if and only if $\rho \circ E|_{\9C_0(Y)}$
  is a state on $\9C_0(Y)$.  By Riesz-Markov theorem
  \cite{ReedSimon80,Conw85}, states on $\9C_0 (Y)$ are uniquely
  represented by regular probability measures on $Y$.  Therefore $E$
  is regular if and only if for any state $\rho$ there exists a unique
  probability measure $m_{E, \rho}$ such that $\rho (E(f))=\int m_{E,
    \rho}(\8dy) \ f(y), \forall f \in \overline {\9C_0} (Y)$.  Since
  the map $\rho\to m_{E,\rho} (\Delta)$ is convex linear in $\rho$, it
  extends uniquely to a linear functional on $\8M_d^*$, i.e. to an
  operator $M_E(\Delta)\in\8M_d$. The map $\Delta \to M_E(\Delta)$,
  uniquely determined by this construction, is clearly a POVM.  Hence,
  $E$ is regular if and only if there exists a POVM $M_E$ such that
  $E(f) = \int_Y M_E(\d y) ~ f(y)$. Of course, $M_E = M_F$ implies $E=
  F$.}

%Conversely, let $M$ be a POVM. For any
%state $\omega \in \8M_d^*$ the relation $\omega_M (\Delta) = \omega
%(M(\Delta))$ defines a probability measure $\omega_M$. The expectation
%of a function $f$ with respect to $\omega_M$, denoted by $\mathbb
%E_{\omega_M}(f)$, defines uniquely a OVE $T_M$ via the relation
%$\mathbb E_{\omega_M} (f)= \omega(T_M(f))$.  Suppose by absurdum that $T_M$ is not regular, namely $\sup\{ T(f) | f\in \9C (Y), 0\le f \le 1\} < \openone_d$. Then there exists a state $\omega$ such that $\|\omega \circ T|_{\9C_0(Y)}\| <1$.    On the other hand, by Riesz-Markov theorem the expectation $\mathbb E_{\omega_M}$ is a state on $\9C_0 (Y)$, leading to the contradiction $1=\| \mathbb
%E_{\omega_M}|_{\9C_0(Y)} \| =\|\omega \circ T_M|_{\9C_0(Y) }\| <1$. Hence, the OVE $T_M$ must be regular.

Theorem \ref{thm-riesz} also provides a characterization of the whole set $\pove Y d$:

\begin{cor}\label{cor:pove-povm}
Let $Y$ be a locally compact Hausdorff space, and let $\bar Y$ be its one-point compactification. Then the following chain of isomorphisms holds 
\begin{equation}
\pove Y d \simeq \pove {\bar Y} d \simeq \9M (\bar Y, d)~.
\end{equation}
\end{cor}
\Proof{Since $\overline {\9C_0} (Y)$ is isomorphic to $\9C(\bar Y)$, one has the
  natural isomorphism $\pove Y d \simeq \pove {\bar Y} d$. Moreover, since $\bar Y$ is compact, one has $\pove {\bar Y} d \equiv \regpove {\bar Y} d$, and, due to Theorem \ref{thm-riesz}, $\regpove {\bar Y} d \simeq \povm {\bar Y} d$.}
%\begin{rem} {\em Theorem \ref{thm-riesz} shows that for compact
%    outcome spaces there is a natural isomorphism between the two
%    convex sets of OVEs and POVMs, since in this case
%    $\pove{Y}{d}\equiv \regpove{Y}{d} \simeq \9M (Y,d)$.  The
%    situation is much different for the non-compact case, where the
%    regular OVEs are a proper subset among all OVEs. According to
%    Corollary \ref{cor:pove-povm}, for noncompact outcome spaces the
%    correspondence between POVMs and OVEs is 
%\begin{equation}
%\9M (Y, d) \simeq
%    \regpove Y d \subset \pove Y d \simeq \9M (\bar Y, d)~.
%\end{equation}
%Clearly, not any POVM in $\9M (\bar Y,d)$ can be restricted to a POVM
%in $\9M(Y,d)$\cite{restriction}: for example a POVM $M \in \9M(\bar
%Y,d)$  that is concentrated in a point outside $Y$---such as the POVM $M(\Delta) = \chi_{\Delta} (\bar y) \openone_d$ with
%$\bar y \in \bar Y \setminus Y$---yields the null operator for any
%measurable subset of $Y$.} 
%\end{rem}

\subsection{Convexity and topology}
 
The sets $\pove Y d$ and $\regpove Y d \simeq \povm Y d$ possess a
natural convex structure, namely the convex combination of two
(regular) OVEs is a (regular) OVE.  Operationally, the convex combination of two
quantum measurements corresponds to a random choice of the
corresponding measurement apparatuses with suitable probabilities.
The extreme OVEs are those which cannot be decomposed into nontrivial
convex combinations:
\begin{defi}
  An OVE $E \in \pove Y d$ is extreme if for any couple of OVEs $E_+, E_- \in
  \pove Y d$ the equality $E= 1/2 (E_+ + E_-)$ implies $E_+=E_-=E$.
\end{defi}
Similarly one can define the extreme regular OVEs. The extreme points
of $\pove Y d$ and $\regpove Y d$ will be denoted by $\partial
 {\pove{Y}{d}}$ and $\partial{\regpove Y d}$, respectively.

The notion of finite convex combination can be generalized to
the notion of barycenter, that includes the possibility of infinite
combinations with arbitrary probability distributions.  For this
generalization, however, one has to first specify a topology on the
set of OVEs.  We will consider here the weak*-topology induced by the
family of seminorms
\begin{equation}
w_{\rho,f}(E)=|\rho (E(f))|
\end{equation}
with $\rho\in \8M_d^*$ and $f\in\overline {\9C_0} (Y)$. This topology has a direct
operational interpretation in quantum mechanics: what can be tested in
experiments are indeed the expectation values $\rho(E(f))$ where
$\rho$ is the state of the quantum system, $E$ describes the
measurement, and $f$ is a function of the outcome.  If the expectation
values $\rho(E_n(f))$ obtained in a sequence of measurements $\{E_n\}$
converge to $\rho(E(f))$ for any state $\rho$ and any function $f$,
then the sequence of measurements $\{E_n\}$ converges to $E$.
Accordingly, the weak*-closure $\overline {\9U}$ of a set of quantum measurements
contains all OVEs that can be arbitrarily approximated with
measurements in $\9 U$ in the sense of expectation values.

Let $\sigma (\pove Y d)$ be the Borel $\sigma$-algebra generated by weak*-open sets.Then we have the following definition: 

% The
%average is $\mathbb E_{\mu} (\eta_{\xi,f}|_{\9U}) = \int_{\9U} \mu(\d
%T) ~\xi(T(f))$.
\begin{defi}
  Let $p$ be a probability distribution on $\sigma( \pove Y d)$ and
  $\9U \in \sigma (\pove Y d)$ be a Borel set. An OVE $E$ is the
  barycenter of $\9U$ with respect to $p$, denoted by
\begin{equation}
E = \int_{\9U} p(\d F)~ F
\end{equation} 
if for any $\rho \in \8M^*_d$ and for any $f \in \overline{\9C}_0 (Y)$ the  following relation holds:
\begin{eqnarray}\label{MeasTheoDefBary}
%&& p(\9U) =1\\ 
  &&\rho(E(f)) =\int_{\9U}  p(\d F)  ~\rho(F(f))~.\label{integ}
\end{eqnarray} 
\end{defi}

Notice that the integral in Eq.  (\ref{integ}) is well defined since
the expectation value $\rho(F(f))$ is by definition a
weakly*-continuous function of $F$, and therefore can be integrated with
respect to any Borel measure $p (\d F)$.

%%%
\section{Characterization of extreme POVMs}
%%%
\label{sec-2}
  
%%%
\subsection{Existence of densities for OVEs in finite dimensions}
%%%
We first prove that every regular OVE admits a density with respect to a
finite measure on $Y$.
%%%
\Lem{density}{For any regular OVE $E\in\regpove{Y}{d}$ there exist a
  regular finite measure $\mu_E$ on $Y$ and a positive density function
  $D_E\in L_\infty(Y, \mu_E)\otimes \8M_d$ such that for any $f \in \overline {\9C_0} (Y)$
\begin{equation}
E(f)=\int \mu_E(\8dy) \ D_E(y)  \ f(y)~.
\end{equation}
 The density function $D_E$ has unit trace,
namely $\tr [D_E(y)] = 1$ $\mu_E$-almost
everywhere.} %Moreover, the support of $T$ is contained in the support of $\hat \mu$.
%%%  
\Proof{Let $\tr$ be the trace on $\8M_d$. Then $\hat \mu_E:=\tr \circ
  E$ is a positive functional with norm $\| {\hat \mu_E}\| =d$.  Since
  $E$ is regular, by Riesz-Markov theorem $\hat \mu_E$ can be
  represented by a regular finite measure $\mu_E$ on $Y$. Moreover,
  the dominance relation $ E\le \hat \mu_E \ \I_d$ holds. Indeed, for
  any positive function $f$ one has $E(f) \le \| E(f) \| \ \I_d \le
  \tr[E(f)] \ \I_d =\hat \mu_E(f) \ \I_d$. The Radon-Nikodym Theorem
  for OVEs [Lemma \ref{Radon-Nikodym} of the Appendix] then guarantees
  the existence of a positive density $D_E \in L_{\infty} (Y, \mu_E)
  \otimes \8M_d$, namely an operator valued function $D_E (y)$
  satisfying the relation $E(f) = \int \mu_E (\d y) \ D_E(y) \ f(y)$.
  Finally, for any $f \in \overline {\9C_0} (Y)$ we have
\begin{eqnarray}
\int \mu_E(\d y) \
  f(y) &=& \hat \mu_E (f)\\
 &=& \tr [E(f)]\\
 &=&\int \mu_E(\d y) \ \tr [D_E (y)]
  \ f(y)~,
\end{eqnarray}
which implies $\tr[D_E(y)] =1$ $\mu_E$-almost everywhere.}

%%%
\subsection{Extreme OVEs}
%%%
We show here that every extreme POVM in dimension $d$ is concentrated
on a finite set of $k \le d^2$ points. This is done by
characterizing the set of extreme regular OVEs.

\Lem{NecCondForExtreme}{Let $E\in \regpove{Y}{d}$ be a regular OVE,
  and let $\mu_E$ be the finite measure associated to $E$ as in Lemma \ref{density}.  If $E$ is extreme, then
  the associated measure $\mu_{E}$ is concentrated on a finite set of
  $k\le d^2$ points.}
 
\Proof{Let $\mu_E$ and $D_E$ be the finite measure and the density
  function associated to $E$ as in Lemma \ref{density}, respectively.
  The density $D_E \in L_{\infty}(Y, \mu_E) \otimes \8M_d$ induces a
  linear operator $\hat D_E \:\8M_d^*\to L_{\infty}(Y,\mu_E)$
  according to $\hat D_E (\rho)=(\id\otimes\rho)D_E$, $\id$ denoting
  the identity map on $L_{\infty} (Y, \mu_E)$. The dimension of the
  image of $\hat D_E$ is clearly bounded by $d^2$, which is the
  dimension of its domain.  By absurdum, suppose that $E$ is extreme
  and the support of the measure $\mu_E$ contains more than $d^2$
  points.  Since the space $Y$ is Hausdorff, this implies that the
  dimension of $L_{\infty} (Y, \mu_E)$ is strictly larger than $d^2$
  \footnote{Indeed, for any finite collection of points $\{y_i \in
    \supp {\mu_E} ~|~ i=1, \dots, k <\infty\}$ there is a collection
    of open neighborhoods $\{U_i~|~ i=1, \dots, k\}$ with $U_i \cap
    U_j = \emptyset$ for $i\not =j$.  If the support contains more
    than $d^2$ points, then the dimension of $L_{\infty} (Y,\mu_E)$ is
    clearly larger than $d^2$, as the indicator functions of the sets
    $U_i$ are linearly independent elements of $L_{\infty} (Y,
    \mu_E)$.  }.  Hence, there is at least one function $h \in
  L_{\infty} (Y, \mu_E)$ that is linearly independent from all
  elements in the image of $\hat D_E$.  The function $h$ can be chosen
  to be real without loss of generality.  Moreover, since $\mu_E$ is a
  finite measure on $Y$, the inclusion $L_{\infty} (Y, \mu_E)
  \subseteq L_2 (Y,\mu_E)$ holds, and $S = \{ \alpha h + \beta \hat
  D_E (\rho)~|~ \alpha, \beta \in \7C, ~\rho \in \8M_d^*\}$ is a $(d^2
  +1)$-dimensional closed subspace of $L_2(Y, \mu_E) \cap L_{\infty}
  (Y, \mu_E)$.  It is then possible to choose a non-zero real function
  $g\in S$ with $\|g\|_{\infty} < \infty$ that is orthogonal to all
  elements in the image of $\hat D_E$, namely
\begin{equation}\label{ortog}
\< g, \hat D_E (\rho)\> = \int_Y \mu_E(\d y) ~ g(y) ~ \hat D_E (\rho) (y) =0~.
\end{equation} 
This implies the decomposition $E=\frac{1}{2}(E_++E_{-})$ where

\begin{equation}
E_{\pm }(f)= E ( (1 \pm \tau g) f)~,  \qquad \tau = \frac 1 {2 \|g\|_{\infty}}~.
\end{equation}
We claim that the above decomposition is a nontrivial convex
decomposition of $E$, in contradiction with the fact that $E$ is
extreme.  First, $E_{\pm}$ is a positive map: $E_{\pm}(f) = E ((1 \pm
\tau g) f)\ge 0$ for any positive function $f\ge 0$. The normalization
$E_{\pm} (\I_Y) =\I_d$ follows from the relation $\rho(E_{\pm}(\I_Y))=
\rho(E(\I_Y)) \pm \tau \langle g, \hat D_E (\rho)\rangle =\rho (\I_d)$
holding for any $\rho \in M_d^*$ due to Eq. (\ref{ortog}). Hence,
$E_{\pm}$ is an OVE. Finally, the decomposition is nontrivial, namely
$E_+ \not = E_-$. Indeed, one has $E_+ (f) - E_{-} (f) = 2 \tau E(fg)$,
which cannot be zero  for any $f\in \overline {\9C_0} (Y)$, otherwise using Lemma \ref{density} one would have also
\begin{eqnarray}
0 = \tr[E(fg)]&=&\int_Y \mu_E (\d y) ~  \tr[ D_E (y)] ~f(y)~ g(y)\\
&=& \int_Y \mu_E (\d y)   ~f(y)~ g(y) = \< g,f\>  
\end{eqnarray} 
for any $f\in \overline {\9C_0} (Y)$, in
contradiction with the fact that $g \in L_{2} (Y, \mu_E)$ is
nonzero by construction. }

As a consequence of the previous Lemma one can reduce the
characterization of extreme OVEs with locally compact Hausdorff space
$Y$ to the characterization of extreme OVEs with finite outcome
space:

\Thm{Characterization of extreme regular OVEs}{thm-extreme} {Let $Y$
  be a locally compact Hausdorff space, and $X$ be a finite set with
  cardinality $|X| =\min\{d^2, |Y| \}$. A regular OVE $E\in
  \regpove{Y}{d}$ is extreme if and only if there exists an extreme
  OVE $P\in \pove{X}{d}$ and an injective function $\varphi \in \9C
  (X,Y)$ such that the following identity holds:
\begin{equation}\label{ExtremePull}
E(f)=P(f\circ\varphi) \qquad \forall f \in \overline {\9C_0} (Y)~.
\end{equation}}
\Proof{Suppose that $E$ is extreme. Then, according to Lemma
\ref{NecCondForExtreme}, the measure $\mu_E$ is concentrated on a
finite set of points $\{y_i~|~ i=1, \dots, k\}$ with $k\le d^2$,
namely $\mu_E(\Delta) = \sum_{i=1}^k \chi_{\Delta} (y_i) ~ p_i $, with
$
p_i\ge 0, \sum_i p_1 =1$.  Using Lemma \ref{density} one obtains
  \begin{eqnarray}
  E(f) &=& \int_Y \mu_E (\d y) ~D_E(y)~  f(y)\\
  &=& \sum_{i=1}^k  p_i    D_E (y_i)  ~f(y_i)\\
  &=& \sum_{i=1}^{|X|}  P_i ~ f(y_i)\\
  &=& P(f\circ \varphi)~,
\end{eqnarray}
where $X=\{1, 2, \dots, \min\{d^2 , |Y| \}$, $P (h) = \sum_i h(i)
P_i$ for any $h \in \9C (X)$,
\begin{equation}
P_i = \left\{ 
\begin{array}{ll} 
p_i D_E (y_i) \qquad &  i =1, \dots, k\\
0  & i=k+1, \dots ,|X|
\end{array}
\right.
\end{equation}
and $\varphi \in \9C (X, Y)$ is any injective function such that $\varphi (i)=y_i,
\forall i =1,\dots, k$.  Obviously $P$ must be extreme in $\pove X
d$, otherwise one would obtain a non-trivial convex decomposition of
$E$.  Conversely, suppose $E$ is as in Eq.  (\ref{ExtremePull}).
Then the measure $\mu_E$ associated to $E$ has finite support $\supp {
  \mu_E} \subseteq \varphi (X) =\{y_i~|~ i=1, \dots, \min\{d^2,|Y| \}\}$.  Suppose
that $E = 1/2 (E_+ + E_-)$ with $E_{\pm} \in \pove Y d$.  Since
$E_{\pm}$ are positive maps, we have $E_{\pm} \le 2 E \le 2 \hat \mu_E
\I_d$, where $\hat \mu_E$ is the functional associated to $\mu_E$.
Due to the Radon-Nikodym theorem for OVEs [Lemma \ref{Radon-Nikodym}
of the Appendix], $E_{\pm}$ admits a density with respect to $\mu_E$,
whence 
\begin{equation}
\begin{split}
E_{\pm} (f) &= \int_Y \mu_E (\d y) ~ D_{\pm} (y) ~ f(y)\\
 &=\sum_{i \in X}  p_i D_{\pm} (y_i) ~ f(y_i)\\ 
 &= P_{\pm} (f \circ \varphi)~. 
\end{split}
\end{equation}
upon defining the OVE $P_{\pm} \in \pove X d$ by $P(h) = \sum_{i \in
  X} p_i D_{\pm}( y_i) h(i), \forall h \in \9C(X)$.  Moreover, since
$Y$ is a locally compact Hausdorff space and $\varphi$ is injective,
the mapping $f \mapsto f \circ \varphi$ is surjective on $ \9C (X)$
\footnote{ Since any locally compact Hausdorff space is completely
  Hausdorff, for any $i \in X$ there exists a function $f_i \in
  \overline {\9C_0} (Y)$ that separates $y_i$ from the finite set
  $\{y_j~|~ j \in X, j \not =i \}$, namely $f_i(y_j)=\delta_{ij}$.  As
  a consequence, $h_i (j) := f_i \circ \varphi (j) = f_i (y_j) =
  \delta_{ij}$. Since the functions $h_i$ are a basis for the finite
  dimensional vector space $\9C(X)$, the map $f \mapsto f \circ
  \varphi$ is surjective.  }.  Therefore we have $P(h) =
1/2 (P_+(h) + P_-(h))$ for any $h \in \9C (X)$, i.e. $P = 1/2 (P_+ +
P_-)$ and, due to extremality of $P$, $P_+ = P_- = P$. In conclusion,
we obtained $E_ + = E_- =E$, i.e.  $E$ is extreme.}

For any continuous function $\varphi : X \to Y$,  we now define the
continuous map $\hat \varphi : \pove X d \to \pove Y d$, which maps $
P \in \pove X d$ to the OVE $\hat \varphi (P) \in \pove Y d$ defined
by the relation
\begin{equation}\label{hatphi}
\hat  \varphi (P)  (f) = P (f \circ \varphi) \qquad  \forall f \in \overline {\9C_0} (Y)~.  
\end{equation}
We denote by $\9I (X, Y)$ the set of injective functions in $\9C(X,
Y)$, and define a map $\iota_{X,Y}$ that transforms subsets of $\pove
X d$ into subsets of $\pove Y d$ as follows
\begin{equation}\label{iota}
  \iota_{X,Y} (C) := \{\hat \varphi (P)~|~ \varphi \in \9I (X,Y), P \in C  \} \qquad \forall C \subseteq \pove X d~.
\end{equation}
With this definition, we can state the following
\Cor{pullext}{
Let $X, Y$ be as in Theorem \ref{thm-extreme}, and let $\bar Y$ be the one point compactification of $Y$.  Then the following equalities hold:
\begin{eqnarray}
\label{pullone}  \partial\regpove Y d &=& \iota_{X,Y} \left(\partial \pove X d \right) \\
 \label{pulltwo} \partial{\pove Y d} &=& \iota_{X,\bar Y} \left(\partial \pove X d \right) ~.
\end{eqnarray}
Moreover, $\partial{\regpove Y d} = \partial{\pove Y d} \cap \regpove
Y d$.}

\Proof{Eq. (\ref{pullone}) directly follows from Theorem
  \ref{thm-extreme}. Eq. (\ref{pulltwo}) follows from Theorem
  \ref{thm-extreme} and from the identification $\pove Y d \simeq
  \pove {\bar Y} d \equiv \regpove {\bar Y} d$. Finally, combining
  Eqs. (\ref{pullone}) and (\ref{pulltwo} we have the inclusion
\begin{equation}
\begin{split}
  \partial \regpove Y d &= \iota_{X, Y} (\partial \pove X d) \subseteq
  \iota_{X, \bar Y} (\partial \pove X d) \cap \regpove Y d \\
  & = \partial \pove Y d \cap \regpove Y d~.
\end{split} 
\end{equation} Conversely, an OVE $E \in
\partial \pove Y d$, given by $E (f) = P (f \circ \varphi)=\sum_i f
(\varphi (i)) P_i$, is regular only if $\varphi (i) \in Y$ for any $i$
such that $P_i \not =0$. Therefore, there exists an injective function
$\tilde \varphi \in \9I(X,Y)$ such that $E (f) = P (f \circ \tilde
\varphi)$, namely $E \in \partial \regpove Y d$.  In conclusion, we
have $\partial \regpove Y d = \partial \pove Y d \cap \regpove Y d$.}

The characterization of extreme POVMs immediately follows as a corollary from the previous Theorem:
\begin{cor}[Extreme POVMs]\label{cor:extPOVM}
  {Let $X$ and $Y$ be as in Theorem \ref{thm-extreme}. A POVM $M \in
    \povm Y d$ is extreme if and only if there exist an injective function $\varphi
    \in \9 C (X,Y)$, and an extreme finite-outcome POVM $P \in \povm
    X d$ such that for any Borel set $\Delta \in \sigma (Y)$
\begin{equation}
M(\Delta) = \sum_{i\in X} \chi_{\Delta} (\varphi(i)) ~ P_i~,
\end{equation}
$\chi_{\Delta}$ denoting the indicator function of $\Delta$.}
% and the map $T_P$ associated to $\{P_i\}$ as in Theorem \ref{finite-ext} is invertible.}
\end{cor}
\begin{rem}
  \emph{The above characterization implies that any extreme quantum
    measurement $M \in \povm Y d$ with locally compact outcome space
    $Y$ can be realized by first performing finite-outcome measurement
    $\{P_i~|~ i\in X\}$, and then, conditionally to outcome $i\in X$,
    by declaring outcome $\varphi(i) \in Y$. In such a scheme the
    function $\varphi \in \9C (X,Y)$ simply represents a classical
    post-processing of the measured data. It is worth stressing that
    for extreme POVMs such a post-processing must be injective:
    $\varphi (i) = \varphi (j)$ only if $i=j$.}
\end{rem}

For the sake of completeness we conclude this Section with a
characterization of extreme OVEs in $\pove X d$, which coincides with
the characterization of extreme finite-outcome POVMs of Ref.
\cite{Stoerm}.

\Thm{Extreme finite-outcome OVEs}{finite-ext}{Let $P \in \pove X d$
  be an OVE with finite outcome space, given by $P(h) = \sum_i h_i
  P_i$, $P_i \in M_d$. Denote by $\2H_i$ the
  range of $P_i$ and by  $\9B (\2H_i)$  the algebra of linear operators on $\2H_i$.  Then, $P$ is extreme if and only if the map $T_P:
  \bigoplus_{i\in X} \9B (\2H_i) \to \8M_d$ given by
\begin{equation}
T_P \left(\bigoplus_i A_i \right) = \sum_{i\in X} \sqrt{P_i} A_i
  \sqrt {P_i}
\end{equation}
is injective.}

\Proof{Suppose $P= 1/2 (P_+ + P_-)$ for some $P_{\pm} \in \pove X d$.
  This implies that $2P - P_{\pm} \ge 0$, i.e. $P_{\pm}$ is dominated
  by $2P$. Let $(\2H_{P}, \pi_{P}, V_{P})$ be the minimal Stinespring
  representation \cite{stine} of $P$, given by $\2H_P = \bigoplus_i \2H_i$,
  $\pi_P(h) = \bigoplus_i h_i \, \I_{\2H_i}$, and $V_P =\sum_i
  \sqrt{P_i} \otimes |i\>$ (here the tensor with $|i\>$ denotes the
  embedding of $\2H_i$ in $\2H_P$ and the operator $\sqrt{P_i} \otimes
  |i\>$ is defined by $(\sqrt{P_i} \otimes |i\>) \varphi = (\sqrt{P_i}
  \varphi) \otimes |i\>$, for any $\varphi \in \7C^{d}$).  The Radon-Nikodym
  theorem for completely positive maps \cite{arveson,belavkin,raginsky} then implies $P_{\pm} (h) =
  V_P^{\dag} D_{\pm} \pi_P(h) V_P$, for some positive operator
  $D_{\pm}$ in the commutant of $\pi_P$, i.e. in $\bigoplus_i \9B (\2
  H_i)$. Accordingly, we have $P_{\pm} (h) = \sum_i h_i \sqrt{P_i}
  D^{\pm}_i \sqrt{P_i}$ with $D^{\pm}_i \in \9B (\2 H_i)$. Since we
  have $P_{\pm} (\I_X) =T_P (D_{\pm})$, the normalization condition
  $P_{\pm} (\I_X) = \I_d$ is satisfied with $P_+ \not = P_-$ if and
  only if the map $T_P$ is not injective, i.e.  $P$ is not extreme
  if and only if $T_P$ is not injective.}

%%%
\section{Topological properties of $\pove Y d$ and $\regpove Y d$}
\label{sec-3}
Operator valued expectations are elements of the Banach space $\cbm
Y d$ of bounded maps from $\overline {\9C_0} (Y)$ to $\8M_d$, which is
naturally isomorphic to the Banach space $\overline {\9C_0} (Y)^* \otimes
\8M_d$: 

\Lem{lemma-1}{Let $V$ denote the Banach space
  $V=\overline{\9C}_0(Y)\otimes\8M_d^*$, equipped with the cross norm
\begin{equation}
\|B\|=\inf\left\{\sum_i \|f_i\| \ \|\rho_i\|_1\biggm| B=\sum_i f_i\otimes\rho_i\right\} \; ,
\end{equation}
$\|\cdot \|_1$ being the norm on $\8M_d^*$. Then, the Banach space $\cbm Y
d$ is isomorphic to the dual Banach space $V^*$.  } \Proof{Any map $E
\in\cbm{Y}{d}$ induces a linear functional $\hat E \in V^*$, which is
defined on product vectors by $\hat E(f\otimes\rho):=\rho( E(f))$ and
uniquely extended on $V$ by linearity. The correspondence $E \mapsto \hat E$
is invertible and preserves the norm, i.e. $\| E\| = \|\hat E\|_{V^*}$
where $\|\hat E\|_{V^*}=\sup_{B,\|B\|=1} |\hat E(B)|$.  Indeed, on the
one hand we have $\|E\|= \sup_{\rho,\|\rho\|_1=1}
\sup_{f,\|f\|=1}|\rho(E(f))| \leq \sup_{B,\|B\|=1} |\hat E(B)|=\|\hat
E\|_{V^*} $. On the other hand, for any possible decomposition of $B
\in V$ as $B=\sum_i f_i\otimes\rho_i$ we have $|\hat E(B)|=\biggm|\sum_i \rho_i(E(f_i))\biggm|\leq \|E\|\sum_i\|\rho_i\|_1\|f_i\|$.
Taking the infimum over all decompositions we get $\|\hat
E\|_{V^*}\leq \|E\|$, and, therefore, $\|E\|=\|\hat E\|_{V^*}$.}

Owing to the above isomorphisms, in the following we identify the map
$E$ with the functional $\hat E$ and the set $\cbm Y d$ with $V^*$.

\Lem{compact}{The convex set $\pove{Y}{d} \subset V^*$ is closed and
  compact in the weak*-topology.}  \Proof{Suppose that a net
  $(E_a)_{a\in A}\subset\pove{Y}{d}$ converges to the linear
  functional $E \in V^*$ in the weak*-topology, i.e.  $\lim_a
  E_a(B)=T(B)$ for any $B\in V$.  In particular, for $B = f \otimes
  \rho$ we have $\rho(E(f)) =\lim_a \rho (E_a (f))$.  Since for any
  positive function $f\ge 0$ one has $E_a (f) \ge 0$ for any $a \in
  A$, one necessarily has also $E(f) \ge 0$.  Similarly, $E_a (\I_Y) =
  \I_d, \forall a \in A$ implies $E(\I_Y) = \I_d$.  This proves that
  $E$ is an element of $\pove{Y}{d}$, whence $\pove{Y}{d}$ is
  weak*-closed. Finally, since $\pove{Y}{d}$ is contained in the unit
  ball of $V^*$ (see Eq. (\ref{intheball})), it is weak*-compact due to the
  Banach-Alaoglu theorem.}

\Lem{metrizable}{If $Y$ is second countable, then the set $\pove Y d$
  is metrizable.}

\Proof{Since $Y$ is second countable, also its one point
  compactification $\bar Y$ is second countable. Being a second
  countable compact space, $\bar Y$ is then metrizable due to Urysohn's
  metrization theorem \cite{urysohn}. This implies that the Banach space of
  continuous functions $\9C (\bar Y)$ is separable \cite{Kreinkrein}.
  Moreover, since the dimension $d$ is finite, the Banach space $V=
  \9C(\bar Y) \otimes \8M^*_d$ is also separable. We now invoke the
  well known result that the unit ball in the dual of a separable
  Banach space is weak*-metrizable \cite{dunford}.  Since $\pove Y d$ is a
  subset of the unit ball in $V^*$, it is metrizable.}

We conclude with the following useful Lemma about the set of regular OVEs

\Lem{lemma-6}{The set $\regpove Y d$ is a $G_{\delta}$-set, namely
  there exists a sequence of open sets $\{U_n\}$ such that $\regpove Y
  d = \bigcap_n U_n$. Moreover, if a regular OVE $E \in \regpove Y d$
  is the barycenter of $\pove Y d$ with respect to a probability
  measure $p_E$, then $\regpove Y d$ has unit measure, i.e. $p_E (\regpove Y d) =1$.}

\Proof{Definition \ref{defi-regpove} of a regular OVE is equivalent to
  the condition
\begin{equation}\label{regocond}
\sup\{ \tau (E(f)) ~|~ f\in\9C_0(Y), 0\leq
  f\leq\I_{Y}\}=1~,
\end{equation}
where $\tau = \tr/d$ is the normalized trace on $\8M_d$. Denote by
$\9S_n \subseteq \pove {Y} d$ the set of OVEs $E \in \pove {Y} d$ such
that
\begin{equation}
\sup\{
\tau (E(f)) ~|~ f\in\9C_0(Y), 0\leq f\leq\I_{Y}\}\le 1- \frac 1 n ~.
\end{equation}
The set $\9S_{n}$ is a weak*-closed subset of $\pove {Y} d$. If an OVE
$E \in \pove {Y} d$ is not regular, then it must be in one of the sets
$\9S_n$ for some $n \in \mathbb N$, namely
\begin{equation}\pove {Y} d \setminus \regpove Y d = \bigcup_{n} \9S_n~.
\end{equation} 
Since $\regpove Y d = \bigcap_n \left ( \pove Y d \setminus
  \9S_n\right)$ and the each set $U_n := \pove Y d \setminus \9S_n$ is
open, $\regpove Y d$ is a $G_{\delta}$-set.  In particular, $\regpove
Y d$ is measurable.  Moreover, for any $f\in\9C_0(Y),0\leq
f\leq\I_{Y}$ we have the following bound
\begin{eqnarray}
\tau (E(f)) &=& \int_{\pove {Y} d} p_E (\d F)~ \tau (F(f))\\
&=& \int_{\9S_n} p_E (\d F) ~\tau (F(f)) + \int_{\pove { Y} d \setminus \9S_n } p_E(\d F)~ \tau (F(f))\\
 &\le& (1-1/n)~ p_E (\9S_n) + (1-p_E (\9S_n))\\ &=& 1 -p_E(\9S_n)/n~.
\end{eqnarray} 
Taking the supremum with respect to $f$ and using the regularity
condition (\ref{regocond}), we then obtain $p_E (\9S_n)=0$ for any
$n$. As a consequence, $\regpove Y d$ has unit measure.}

%%%
\section{Barycentric decomposition }\label{sec-4}

\subsection{Case of second countable outcome spaces}

According to Lemmas \ref{compact} and $\ref{metrizable}$, the set $\pove Y d$ is compact metrizable set.  Choquet's theorem \cite{Choquet,BishopDeLeeuw} then implies the following:

\Lem{lemma-5}{Let $Y$ be a second countable locally compact Hausdorff
  space. Any OVE $E \in \pove{Y}{d}$ is the barycenter of
  $\partial \pove Y d$ with respect to a suitable probability
  measure $p_E$.}

\Proof{Direct application of Choquet's theorem.}

We now combine the Choquet representation with the regularity
condition:

\Thm{Barycentric representation of regular OVEs}{thm-main}{Let $Y$ be a
  locally compact second countable Hausdorff space. Then, any regular OVE
  $E\in\regpove{Y}{d}$ is the barycenter of the set $\partial \regpove Y d$ with
  respect to a probability distribution $p_E$.}
%\begin{eqnarray}
%&&\mu_T (\9F (Y, d)) =1 \\
%&&\rho(T(f)) = \int_{\pove Y d}   \mu_T (\d E)   ~\rho (E(f)) \qquad \forall \rho \in M_d^*, \forall f \in \overline {\9C_0} (Y)~.
%\end{eqnarray}

\Proof{%The proof exploits the isomorphism $\pove Y d \simeq \pove
 % {\bar Y} d$, where $\bar Y$ is the one point compactification of
 % $Y$. 
  By Lemma~\ref{lemma-5} any OVE $E\in \pove Y d$ is the barycenter of
  the set $\partial \pove Y d $ with respect to a probability measure
  $p_E$. On the other hand, since $E$ is regular, Lemma \ref{lemma-6}
  requires the set $\regpove Y d$ to have unit measure. Finally, by
  Corollary \ref{pullext} we have $\partial \regpove Y d =
  \partial \pove Y d \cap \regpove Y d$. Since both $\partial \pove Y
  d $ and $\regpove Y d$ are measurable sets with unit measure, also
  their intersection enjoys this property. }

Owing to the affine bijection established by Theorem \ref{thm-riesz},
the present result can be readily translated into a Choquet representation of  POVMs in finite dimensional Hilbert spaces.

\begin{cor}[Barycentric representation of POVMs] Let $Y$ be a locally compact second countable Hausdorff
  space.  Then, any POVM $M \in \9M (Y, d)$ is the barycenter of the
  set $\partial \povm Y d $ with respect to a probability distribution
  $p_M$, namely
  \begin{equation}
    M (\Delta) = \int_{\partial \9M (Y, d) } p_M (\d P) ~  P (\Delta) \qquad \forall \Delta \in \sigma (Y)
  \end{equation}
\end{cor}

\begin{rem} {\em The above Choquet representation, once combined with
    the characterization of extreme POVMs of Corollary
    \ref{cor:extPOVM}, shows that quantum measurements with
    second-countable outcome space can always be interpreted as
    randomizations of extreme finite-outcome measurements,
    corresponding to operator valued measures concentrated on $k\le
    d^2$ points. It is worth stressing that essentially all outcome
    spaces that are relevant for applications in Quantum Mechanics are
    separable and metrizable, and that for locally compact Hausdorff
    spaces these two conditions are equivalent to second countability,
    due to Urysohn's metrization theorem. }
\end{rem}

\subsection{General case}

If the outcome space $Y$ is not second countable, the set $\pove Y d$
is generally not metrizable. In this situation, Choquet's theorem
cannot be applied, and a barycentric decomposition only in terms
extreme points might not be possible.  However, since the
set $\pove Y d$ is compact in the weak*-topology (Lemma
\ref{compact}), we can still exploit Krein-Milman theorem, thus getting the following

\Lem{KM}{Let $Y$ be a locally compact Hausdorff space, and $\overline
  {\partial \pove Y d}$ be the weak*-closure of $\partial \pove Y d$.
  Any OVE $E \in \pove Y d$ is the barycenter of the set $\overline
  {\partial \pove Y d}$ with respect to a probability measure $p_E$.}

\Proof{Direct consequence of Krein-Milman theorem [Lemma \ref{AltKM} of the Appendix].} 

\begin{rem}{\em Notice that in most situations the set $\partial \pove
    Y d$ is not weak*-closed.  For example, take $d=2$ and $Y \equiv X
    =\{1,2,3,4\}$, and consider the OVEs $E_{a}$ defined by $E_a(h) =
    \sum_i h_i E_{i,a}, ~ \forall h \in \9C (X) $ with
    \begin{equation} \begin{split} E_{1,a} =& (\I + \cos a~ \sigma_x +
        \sin a~
        \sigma_y)/4\\
        E_{2,a} =& (\I + \cos a ~\sigma_x - \sin a~
        \sigma_y)/4\\
        E_{3,a} =& (\I - \cos a ~\sigma_x + \sin a~
        \sigma_z)/4 \\
        E_{4,a} = & (\I - \cos a ~\sigma_x - \sin a~ \sigma_z)/4~,
      \end{split} \end{equation} where $\sigma_x = \begin{pmatrix} 0
      &1 \\ 1 &0 \end{pmatrix}, ~\sigma_y = \begin{pmatrix} 0 &-i \\ i
      & \phantom{-}0 \end{pmatrix}, ~\sigma_z = \begin{pmatrix} 1 &
      \phantom{-}0 \\ 0 &-1 \end{pmatrix}$.  Using Theorem
    \ref{finite-ext} it is immediate to verify that the OVE $E_a$ is
    extreme for any $a \in (0,\pi/4]$, while the limit $E = \lim_{a
      \to 0} E_a$ is not extreme, namely $\partial{\pove Y d}$ is not
    closed, whence the decomposition of Lemma \ref{KM} necessarily
    involves some non-extreme OVEs.}  \end{rem} \Thm{Barycentric
  decomposition of regular OVEs}{thm-bary}{Let $Y$ be a locally
  compact Hausdorff space, and $\9F (Y, d)$ be the intersection
  \begin{equation}\label{effe} \9F(Y, d) = \overline {\partial \pove Y
      d} \cap \regpove Y d~.  \end{equation} Then, any regular OVE $E
  \in \regpove Y d$ is the barycenter of the set $\9 F (Y, d)$ with
  respect to a suitable probability measure $p_E$.}  \Proof{By Lemma
  \ref{KM}, any OVE $E$ is the barycenter of the set $\overline
  {\partial \pove Y d}$ with respect to a probability measure $p_E$.
  Combining this fact with Lemma \ref{lemma-6} we immediately obtain
  the thesis.}

Although the set $\9F (Y, d)$ contains also OVEs that
are not extreme, it is simple to realize that it only contains OVEs
that correspond to POVMs concentrated on a finite set of points of
$Y$.  We now conclude the paper by proving this fact, by first showing
that all OVEs in $\overline {\partial \pove Y d}$ correspond to POVMs
concentrated on a finite set of points of $\bar Y$, and then using the
regularity condition.  Let us identify $\9C (X,Y)$ with $X \times Y
\subseteq X \times \bar Y$ and equip it with the product topology.
Accordingly, $\overline{\9I (X,Y)}$ denotes the closure of the set of
injective functions in $\9C (X, Y)$.  Define the map $\bar \iota_{X, Y
}$ transforming subsets of $\pove X d$ into subsets of $\pove X d$ as
follows \begin{equation}\label{bariota} \bar \iota_{X, Y} ( C) := \{
  \hat \varphi (P)~|~ \varphi \in \overline{\9I (X,Y)}, P \in C \}~,
\end{equation} where the map $\hat \varphi$ is defined as in Eq.
(\ref{hatphi}).  We then have the following: \Lem{aux}{Let $X$ and $Y$
  be as in Theorem \ref{thm-extreme}, and $\iota_{X, Y}$ and $\bar
  \iota_{X, Y}$ be the maps defined in Eqs. (\ref{iota}) and
  (\ref{bariota}), respectively. Then, for any subset $C \subseteq
  \pove X d$, one has \begin{equation} \overline{\iota_{X, Y} (C)}=
    \bar \iota_{X, Y} (\bar C)~.  \end{equation}} \Proof{Let $E$ be a
  point of $\overline{\iota_{X,Y} (C)}$, and take a net $(E_a)_{a\in
    A}\subset \iota_{X,Y} (C)$ converging to $E$.  Since $E_a \in
  \iota_{X,Y} (C) $, one has $ E_a (f) = P_a (f \circ \varphi_a)$,
  with $P_a \in C$ and $\varphi_a \in \9I(X, Y)$.  Moreover, since
  $\overline C$ is compact, the net $(P_a)_{a \in A} \subset \overline
  C$ will have a cluster point $P \in \overline C$.  Similarly, the
  net $(\varphi_{a})_{a \in A} \subset \overline {\9I (X,Y)}$ will
  have a cluster point $\varphi \in \overline{ \9I (X, Y)}$.  We can
  then choose a subnet $(E_b)_{b \in B}$ such that $\lim_{b} P_b = P$
  and $\lim_{b} \varphi_b = \varphi $, thus obtaining \begin{equation}
    E(f) = \lim_b E_b (f) =\lim_{b} P_b (f\circ \varphi_b)= P (f \circ
    \varphi )~.  \end{equation} To evaluate the limit we used the fact
  that $\pove X d$ is finite dimensional, whence the weak*-convergence
  of the net $(P_b)_{b\in B}$ is equivalent to  norm convergence.
  The above equation proves that $E$ is in $\bar \iota_{X, Y} (\bar
  C)$, namely $\overline{\iota_{X,Y} (C) } \subseteq \bar \iota_{X, Y}
  (\bar C)$.  Conversely, let $E$ be a point in $\bar {\iota}_{X,Y}
  (\bar C)$, defined by $E (f) = P (f \circ \varphi)$, with $ P \in
  \bar C$ and $\varphi \in \overline {\9I (X, Y)}$. Take a net
  $(P_{a})_{a\in A} \subseteq C$ such that $\lim_{a} P_a= P$ and a net
  of injective functions $(\varphi_b)_{b \in B} \subseteq X \times Y$
  such that $\lim_{b} \varphi_b = \varphi$. Let us equip $A \times B$
  with the product order, and define the net $E_{a,b} \in \iota_{X, Y}
  (C)$ by $E_{a,b} (f) := P_a (f \circ \varphi_b)$.  Clearly, the net
  $(E_{a,b})_{(a,b) \in A \times B}$ converges to $E$, whence $E \in
  \overline {\iota_{X,Y} (C)}$.  This proves that $\bar \iota_{X, Y}
  (\bar C) \subseteq \overline{\iota_{X, Y} (C)}$.}

As a consequence, we have the following
characterization:

\Lem{lemma-4}{The closure of the set $\partial \pove Y d$ is given by
\begin{equation}
  \overline {\partial \pove Y d} = \bar \iota_{X, \bar Y} \left ( \overline{\partial \pove X d}\right)~,
\end{equation}
namely every $E \in \overline{\partial \pove Y d}$ is of the form
\begin{equation}\label{basta}
E (f) = P (f \circ \varphi) \qquad \forall f \in \9C_0 (Y)
\end{equation}
for some suitable OVE $P \in \pove X d$ and some suitable function
$\varphi \in \9C (X, \bar Y)$, obtained as a limit of injective
functions. }

\Proof{ By Corollary \ref{pullext} we have $\partial \pove Y d = \iota_{X, \bar Y} (\partial \pove X d)$. Application of Lemma \ref{aux} then yields the thesis.}

\Thm{Structure of the set $\9F (Y, d)$}{F}{Let 
\begin{equation}\9K (X, Y) = \overline
  {\9I (X,Y)} \cap \9C (X,Y)
\end{equation}
be the set of continuous functions from $X$ to $Y$ that are limits of
injective functions. Then, the set $\9F (Y,d)$ defined in Eq.
(\ref{effe}) is given by
\begin{equation}
\begin{split}
  \9F (Y ,d) = \left \{ E \in \pove Y d ~|~ E (f) = P(f \circ \varphi ), \phantom{\sum_{i=1}^{d^2} } ~ \right.\phantom{aaaaaaa} & \\
  \left.  \phantom{\sum_i^{d^2}} \varphi \in  \9K (X, Y) ,~
    P \in \overline{\partial \pove X d} \right \} &  
\end{split}
\end{equation}}

\Proof{By definition, $\9F (Y, d) = \overline {\partial \pove Y d}
  \cap \regpove Y d$. On the other hand, by Lemma \ref{lemma-4} an OVE
  $E$ is in $\overline{\partial \pove Y d}$ iff if has the form 
\begin{equation}\label{sonomorto}
E(f)
  = P(f \circ \varphi)= \sum_{i \in X} P_i ~ f(\varphi (i))~,
\end{equation} with $P \in \overline{\partial \pove X d}$
and $\varphi \in \overline{\9I (X, \bar Y)}$.  Clearly, an OVE
$E$ in $\overline{\partial \pove Y d}$ is regular iff the function $\varphi$ in Eq. (\ref{sonomorto}) satisfies $\varphi (X)
\subseteq Y$, namely, iff $\varphi \in  \overline {\9I (X, \bar Y)} \cap \9C (X, Y)$.  We now claim that  $ \overline {\9I (X,\bar  Y)} \cap \9C (X, Y) \equiv \9K (X, Y)$.  Indeed, we have the inclusion $\9K (X, Y) =  \overline {\9I (X, Y)} \cap \9C (X, Y) \subseteq  \overline {\9I (X, \bar Y)} \cap \9C (X, Y)$.
Viceversa, let  $\varphi$ be in  $\overline {\9I (X, \bar Y)} \cap \9C (X, Y) $ and $(\varphi_a)_{a \in A}
\subseteq \9I (X, \bar Y)$ be a net of injective functions such that
$\lim_a \varphi_a = \varphi$.  Since the topology of $\9C (X, \bar
Y) \simeq X \times \bar Y$ contains the topology of $\9C(X, Y) =X
\times Y$, for any neighborhood $U \subseteq \9C (X, Y)$ of
$\varphi$ we have that the net $(\varphi_a)_{a \in A}$ must
eventually be in $U$. Hence, $\varphi$ is the limit of a net of
injective functions in $\9I (X,Y)$ as well.  Therefore, we have
$\varphi \in \overline{\9I (X, Y)} \cap \9C(X, Y) =\9K(X, Y)$, thus proving the reverse inclusion.}

Any OVE in $\9F (Y, d)$ corresponds to a POVM concentrated on $|X| \le d^2$ points of $Y$. Indeed, we have
\begin{equation}
E (f) = P (f \circ \varphi) = \sum_{i=1}^{|X|}  f (\varphi(i))~ P_i  = \int_{Y}  M(\d y) ~ f (y)~,  
\end{equation}
where $M$ is the POVM defined by $M (\Delta) := \sum_{i =1}^{d^2}
\chi_{\Delta} (y_i)~ P_i$ for any Borel set $\Delta$.  The barycentric
decomposition for POVMs is the given by the following:

\Cor{finale}{Let $Y$ be a locally compact Hausdorff space, and let $\9Q (Y,
  d)$ be the subset of $\povm Y d$ defined by
\begin{equation}
\begin{split}
  \9Q (Y ,d) = \left \{ M \in \povm Y d ~|~ M (\Delta) =
    \sum_{i=1}^{d^2}  \chi_{\Delta} (\varphi (i)) ~ P_i ,~ \right.\phantom{aaaa} & \\
  \left.  \phantom{\sum_i^{d^2}} \varphi \in \9K (X, Y) ,~
    P \in \overline{\partial \povm X d} \right \} &
\end{split}
\end{equation} 
Then, any POVM $M \in \povm Y d$ is the barycenter of the set $\9Q (Y,
d)$ with respect to a probability distribution $p_M$, namely,
\begin{equation}
M (\Delta )= \int_{\9Q (Y , d) } p_M(\d P) ~ P(\Delta)  ~, 
\end{equation}
for any Borel set $\Delta$.}

The barycentric representation of POVMs with locally compact Hausdorff
space allows one to interpret quantum measurements on finite
dimensional systems as randomizations of measurements with $k\le d^2$
outcomes, thus providing a rigorous proof of the fact that in finite
dimensions continuous spectrum is equivalent to continuous classical randomness controlling the choice of the measuring apparatus.

\bigskip

%\end{remark}
%%%

{\bf Acknoledgements.} GC gratefully acknowledges financial support from the European Community through the project CORNER.  

\section{Appendix}
\begin{appendix}

  For completeness of the presentation, in the following we provide
  the proofs of two standard results, the former on the existence of
  densities for OVEs and the latter on barycentric decompositions in
  locally convex spaces.

\section{Radon-Nikodym theorem for OVEs}
%%%

The following Radon-Nikodym theorem for OVEs is equivalent to the existence of a density for POVMs in finite dimensions, which in turn is a consequence of the Radon-Nikodym theorem for quantum instruments \cite{DaviesBook,Ozawa,HolevoRadonNiko}.

\Lem{Radon-Nikodym}{ Let $\mu$ be a finite regular measure on $Y$ and let $T\in \pove Y
d$ be an OVE satisfying the dominance condition $T \le \hat \mu \I$,
$\hat \mu \in \overline {\9C_0} (Y)^*$ being the positive functional
associated to $\mu$.  Then, there exists a unique positive operator
density $D\in L_{\infty} (Y, \mu) \otimes \8M_d$ such that
\begin{equation}
T(f) = \int \mu(\d y)~  f (y) ~D(y)~.   
\end{equation}}

\Proof{Since $\hat \mu$ is a positive functional, $S = \hat \mu \I$ is
  a completely positive (CP) map.  Moreover, due to the dominance
  condition, $S-T$ is also a CP-map. The Radon-Nikodym Theorem for
  CP-maps \cite{arveson,belavkin,raginsky} then implies that $T(f) = V_S^*
  \pi_S (f) D V_S$, where $(\2H_S,\pi_S,V_S)$ is the minimal
  Stinespring representation of $S$, and $D$ is a unique positive
  operator in the commutant of $\pi_S (\overline {\9C_0} (Y))$. The
  minimal Stinespring representation of $S$ is easily obtained here by
  the GNS representation of $\hat \mu$, given by $(\2H_{\hat \mu},
  \pi_{\hat \mu}, \Omega_{\hat \mu})$. Indeed, the Hilbert space
  $\2H_S$ can be identified with $\2H_{\hat \mu} \otimes \mathbb
  C^{d}$, the representation $\pi_S$ with $\pi_{\hat \mu} \otimes \I_d$,
  and the isometry $V_S$ is defined by
\begin{equation} 
V_S \psi = \Omega_{\hat \mu} \otimes \psi \qquad \forall \psi \in \mathbb C~.
\end{equation}
Therefore, we have
\begin{equation}
\begin{split}
\< \psi_1, T(f) \psi_2\> &= 
\< \psi_1 , V^*_S \pi_S (f) D V_S \psi_2\>\\
&= \<\Omega_{\hat \mu} \otimes \psi_1,   D~(\pi_{\hat \mu} (f)\otimes \I_d)~ \Omega_{\hat \mu} \otimes \psi_2\> \qquad \forall \psi_1,\psi_2 \in \mathbb C^d~.   
\end{split}
\end{equation}
Finally, the GNS Hilbert space $\2H_{\hat \mu}$ can be identified with
$L_2 (Y, \mu)$, where $\Omega_{\hat \mu}$ is the constant function,
and $\pi_{\hat \mu}$ represents the function $f\in \overline {\9C_0}
(Y)$ by a multiplication operator.  With this identification, the
commutant of $\pi_{\hat \mu} (\overline {\9C_0} (Y)) \otimes \I$ is
$L_{\infty} (Y, \hat \mu)\otimes \8M_d$ \footnote{Due to the
  identification $\overline {\9C}_0 (Y) \simeq \9C (\bar Y)$, the
  commutant of $\pi_{\hat \mu}$ coincides with $L_{\infty} (\bar Y,
  \mu)$. On the other hand, since $\hat \mu$ is regular one has
  $L_{\infty} (\bar Y, \mu) \equiv L_{\infty} (Y, \mu)$.}.  Therefore,
the positive operator $D$ is an operator valued function, yielding
\begin{equation}
\< \psi_1, T(f) \psi_2\> = \int \mu(\d y)~ \<\psi,   D(y)  \psi_2\>~ f(y) \qquad \forall \psi_1,\psi_2 \in \mathbb C^d~.   
\end{equation}
which implies the identity $T(f)= \int \mu(\d y)~ D(y) ~f(y)$.}

\section{Barycentric decomposition from Krein-Milman Theorem}

\Lem{AltKM}{Let $K$ be a compact subset of a locally convex vector
  space $X$. Denote with $\overline{\partial K}$ the closure of
  $\partial K$.  Then, any point $x \in K$ is the barycenter of
  $\overline{\partial K}$ with respect to a suitable probability
  measure $p_x$, namely the relation
\begin{equation}
f(x) = \int_{\overline {\partial K}}  \ p_x (\d E) \ f(E)
\end{equation} 
holds for any function $f\in \9C (K)$.}

\Proof{By Krein-Milman Theorem \cite{Roy88}, any $x\in K$ is in the
  closure of the convex hull of ${\partial K}$, i.e. that there exists
  a net $(x_a)_{a}$ contained in the convex hull such that $\lim_{a}
  x_a =x$. Equivalently, $f(x_a) = \sum_{i} p_i^{(a)} f (x_i^{(a)}) :=
  \hat p^{(a)} (f) $ for any $f \in \9C(K)$, where $\{p_i^{(a)} \}$
  are probabilities and $\{x^{(a)}_i \}$ is a finite set of points in
  $\partial K$.  Clearly, the restriction of the functional $\hat p_a$
  to the C*-algebra $\9C(\overline{\partial K})$ is a state, i.e. a
  positive normalized functional.  Since the set of states is compact,
  the net $(\hat p_a)_{a \in A}$ must have a cluster point $p_x$
  within it.  We then have $f(x) = \lim_a f(x_a) =\lim_{a } \hat p_a (f) = \hat
  p_x (f) = \int_{\overline {\partial K}} p_x (\d E) ~ f(E)$, $p_x$
  being the probability distribution on $\overline{\partial K}$
  associated to $\hat p_x$ by Riesz-Markov theorem.  }

\end{appendix}

\end{document}